\documentclass[preprint,12pt]{elsarticle}

\usepackage{amssymb}
\usepackage{subfig}
\usepackage{graphicx}
\usepackage[colorlinks=true,linkcolor=blue,citecolor=blue]{hyperref}
\usepackage{color}
\usepackage{amsmath,empheq}
\usepackage{amsfonts}
\usepackage{amsthm}
\usepackage{soul}
\usepackage{mathrsfs}
\biboptions{sort&compress}
\makeatletter
\def\ps@pprintTitle{%
  \let\@oddhead\@empty
  \let\@evenhead\@empty
  \let\@oddfoot\@empty
  \let\@evenfoot\@oddfoot
}
\makeatother

	\newcommand{\ncd}{\newcommand}
	\ncd{\mrm}    {\mathrm}
	\ncd{\beq} {\begin{equation}}
	\ncd{\eeq} {\end{equation}}

	\def\d{{\rm d}}

	\def\H{{\mathscr{H}}}
	\def\I{{\mathscr{I}}}
	\def\G{{\mathscr{G}}}

\begin{document}

\begin{frontmatter}

	\title{Contact Hamiltonian Mechanics}

	\author[rvt]{Alessandro Bravetti}
	\ead{alessandro.bravetti@iimas.unam.mx}
	
	\author[local]{Hans Cruz}
	\ead{hans@ciencias.unam.mx}

	\author[saya]{Diego Tapias}
	\ead{diego.tapias@nucleares.unam.mx}

	\address[rvt]{Instituto de Investigaciones en Matem\'aticas Aplicadas y en Sistemas, Universidad Nacional Aut\'onoma de M\'exico, A. P. 70543, M\'exico, DF 04510, M\'exico.}
	
	\address[local]{Instituto de Ciencias Nucleares, Universidad Nacional Aut\'onoma de M\'exico,\\ A. P. 70543, M\'exico, DF 04510, M\'exico.}

	\address[saya]{Facultad de Ciencias, Universidad Nacional Aut\'onoma de M\'exico,\\ A.P. 70543, M\'exico, DF 	04510, Mexico.}

\begin{abstract}
	In this work we introduce contact Hamiltonian mechanics, an extension of symplectic Hamiltonian mechanics, and show that it is a natural candidate for a geometric description of non-dissipative and dissipative systems. For this purpose we review in detail the major features of standard symplectic Hamiltonian dynamics and show that all of them can be generalized to the contact case. 
\end{abstract}


\end{frontmatter}


\newpage\tableofcontents

\section{Introduction}

The Hamiltonian formulation of classical mechanics is a very useful tool for the description of mechanical systems due to its remarkable geometrical properties, 
and because it provides a natural way to extend the classical theory to the quantum context by means of standard quantization. 
However, this formulation exclusively describes isolated systems with reversible dynamics, 
while real systems are constantly in interaction with an environment that introduces the phenomena of dissipation and irreversibility. 
Therefore a major question is whether it is possible to construct a classical mechanical theory that not only contains all the advantages 
of the Hamiltonian formalism, but also takes into account the effects of the environment on the system.

%
Several programmes have been proposed for this purpose (see e.g.~\cite{razavy2005classical} for a recent review).
For example, one can introduce \emph{stochastic dynamics} to model the effect of fluctuations due to the environment on the system of interest.
This leads to stochastic equations of the Langevin or Fokker-Planck type with diffusion terms~\cite{weiss1999quantum,chandrasekhar1943stochastic}. 
A different although related approach is the \emph{system-plus-reservoir} technique, in which the system of interest is coupled
to an environment  (usually modeled as a collection of harmonic oscillators). The system and the environment together are considered as an isolated Hamiltonian system
and after averaging out the environmental degrees of freedom one obtains the equations of motion for the system of interest, including dissipative terms.
This is the case for example of the Caldeira-Laggett formalism~\cite{van1992stochastic,caldeira1981influence,caldeira1983quantum}.
An alternative approach is to propose \emph{effective Hamiltonians} with an explicit time dependence 
that reproduce the correct Newtonian equation, including the dissipative forces. A famous example is the Caldirola-Kanai
(CK) model~\cite{caldirola,kanai,lakshmanan}. 
Another proposal based on a nonconservative action principle,  allows for time-irreversible processes, such as dissipation, to be included at the level of the action  \cite{GalleyPRL, galley2014principle}. 
Finally, a more geometrical attempt towards the description of dissipative systems is given by the so-called \emph{bracket formulation} of dynamical systems~\cite{morrison2009thoughts}.
Here one 
generalizes the standard Poisson bracket to a noncanonical Poisson bracket and exploits the algebraic properties of the latter to include dissipation. 
{The literature on all these proposals is very extensive and it is not our purpose here to review them in detail. We refer the interested reader to the standard references cited above
and references therein.}

Here we discuss a new proposal which consists in extending the symplectic phase space of classical mechanics 
by adding an extra dimension, thus dealing with a contact manifold instead of a symplectic one.
Contact geometry arises naturally in mechanics.
First of all, in describing mechanical systems where the Hamiltonian function explicitly depends on time, one usually appeals 
to an extended phase space, the additional dimension being time,
endowed with the Poincar\'e-Cartan $1$-form, which defines a contact structure on the extended space~\cite{abraham1978foundations,Arnold,goldstein2002classical}.
Besides, the time-dependent Hamilton-Jacobi theory is naturally formulated in this extended phase space~\cite{rajeev2008quantization,rajeev2008hamilton}.
Furthermore, it has recently been argued that symmetries of the contact phase space can be relevant for a (non-canonical) quantization of nonlinear systems~\cite{aldaya2014contact}.

In this work we consider the phase space of any  {(time-independent)} mechanical system (either non-dissipative or dissipative) to be a contact manifold, but we
 take a different route from previous works.
In fact, there are two main differences between our proposal and the previous ones. 
First, we do not assume that the additional dimension is time, letting the additional dimension be represented by 
a non-trivial dynamical variable. 
Second, we derive the equations of motion for the system from \emph{contact Hamiltonian dynamics}, which is the most natural extension of symplectic Hamiltonian dynamics~\cite{Arnold}.

Contact Hamiltonian dynamics has been used already 
in thermodynamics (both equilibrium and not~\cite{Mrugaa:2000aa,favache2010entropy,dolfin2011geometric,Bravetti2015377,shinitiro1,shinitiro2}) 
and in the description of dissipative systems at the mesoscopic level~\cite{grmela2014contact}. 
Furthermore, it has been recently introduced in the study of mechanical systems  
exchanging energy with a reservoir~\cite{bravetti2015liouville,PhysRevE.93.022139}.
However, a detailed analysis of the dynamics of mechanical systems and a thorough investigation of the analogy with standard symplectic mechanics have never been pursued before.
We show that the advantages of contact Hamiltonian mechanics are that it includes within the same formalism both non-dissipative and dissipative systems, giving a precise
prescription to distinguish between them, that it extends canonical transformations to contact transformations,  {thus offering more techniques to find the invariants of motion
and to solve the dynamics,}
and that it leads to a contact version of the Hamilton-Jacobi equation. We argue that these additional properties play a similar role as their symplectic counterparts for dissipative systems.


The structure of the paper is as follows: in section~\ref{section-2}, in order to make the paper self-contained, 
we review the main aspects of the standard mechanics of non-dissipative systems, with emphasis on the symplectic
geometry of the phase space and the Hamilton-Jacobi formulation. In section~\ref{section-3} the same analysis is extended to the case of contact Hamiltonian
systems and it is shown by some general examples that this formulation reproduces the correct equations of motion for mechanical systems with dissipative terms.
 {Besides, an illustrative example (the damped parametric oscillator) is worked out in detail in this section in order to show the usefulness of our method.}
\st{Finally} Section~\ref{section-conclusions} is devoted to a summary of the results and to highlight future directions. In particular, 
we discuss a possible extension of our formalism to quantum systems.  {Finally, in \ref{appendix-A} and \ref{appendix-B} we provide respectively a derivation of the invariants
of the damped parametric oscillator and a constructive proof of the equivalence between the contact Hamilton-Jacobi equation and the contact Hamiltonian dynamics.}

Before starting, let us fix a few important notations that are used throughout the text.
Both symplectic mechanics of conservative systems and contact mechanics of dissipative systems are presented first in a coordinate-free manner and then
in special local coordinates -- canonical and contact coordinates -- labelled as $(q^{a},p_{a})$ and $(q^{a},p_{a},S)$ respectively. 
Moreover, the symplectic phase space is always indicated by $\Gamma$, while the contact phase space by $\mathcal T$. 
The extension of any geometric object
to a quantity that explicitly includes time as an independent variable is always indicated with a superscript E over the corresponding object (e.g.~$\Gamma^{\tiny\mbox{E}}$).
Finally, we always use the notation $H$ for the usual symplectic Hamiltonian function 
and $\H$ for the corresponding contact analogue.


\section{Symplectic mechanics of non-dissipative systems}
\label{section-2}
The description of isolated mechanical systems can be given in terms of the Hamiltonian function and of 
Hamilton's equations of motion in the phase space, which has a natural symplectic structure.
In this section we review Hamiltonian dynamics in the symplectic phase space, in order to compare it with the generalization to the contact phase space that is given in the next section.

\subsection{Time-independent Hamiltonian mechanics}\label{sectionTindsympl}
The phase space of a conservative system is the cotangent bundle of the configuration manifold, which is a $2n$-dimensional manifold $\Gamma$. 
Such manifold is naturally endowed with a canonical $1$-form $\alpha$, 
whose exterior derivative $\Omega=\d\alpha$ is non-degenerate, and therefore defines the standard symplectic form on $\Gamma$. 
Given a Hamiltonian function $H$ on $\Gamma$, Hamilton's equations of motion follow from 
	\beq\label{Hameq1}
	-\d H = \Omega(X_{H})\,,
	\eeq 
with $X_{H}$ the \emph{Hamiltonian vector field} defining the evolution of the system.
By a theorem of Darboux, one can always find local coordinates $(q^{a},p_{a})$ with 
$a=1,\dots,n$ -- called \emph{canonical coordinates} -- in which the canonical form is expressed as
	\beq\label{taut}
	\alpha=p_{a}\d q^{a}\,,
	\eeq
{where here} and in the following Einstein's summation convention over repeated indices is assumed.
In such coordinates
	\beq\label{Omega}
	\Omega=\d\alpha=\d p_{a}\wedge \d q^{a}
	\eeq
and from \eqref{Hameq1} it follows that the Hamiltonian vector field reads
	\beq\label{XH}
	X_{H}=
		- \frac{\partial {H}}{\partial q^a}\frac{\partial }{\partial p_a} 
		+ \frac{\partial {H} }{\partial p_a} \frac{\partial }{\partial q^a}\,.
	\eeq
Usually, the canonical coordinates $q^{a}$ and $p_{a}$ correspond to  the particles' generalized positions and momenta. 
From \eqref{XH} the equations of motion take the standard Hamiltonian form
	\beq\label{Hameq2}
	\dot q^{a}=\frac{\partial H}{\partial p_{a}}\, , \qquad \dot p_{a}=-\frac{\partial H}{\partial q^{a}}\,.
	\eeq
A system whose evolution is governed by \eqref{Hameq2} is usually called a \emph{Hamiltonian system}.
 { The time evolution of any (not explicitly time-dependent) function $G \in C^{\infty}(\Gamma)$ 
is determined by the phase space trajectories generated 
by the Hamiltonian vector field $X_H$, that is  
	\beq\label{LiouvilleEq0}
	\frac{\d G}{\d t}  =   {X_H [G]}=\Omega(X_{H},X_{G})=\{G,H\}_{(q^{a},p_{a})}\,,
	\eeq
where we have introduced the notation $\{G,H\}_{(q^{a},p_{a})}$ for the standard \emph{Poisson bracket} between the two functions $G$ and $H$, 
which in canonical coordinates reads
	\beq\label{LiouvilleEq}
	\{G,H\}_{(q^{a},p_{a})}  = \frac{\partial {G}}{\partial q^a}\frac{\partial {H}}{\partial p_a}-\frac{\partial {G}}{\partial p_a}\frac{\partial {H}}{\partial q^a}\,.
	\eeq
Equations \eqref{LiouvilleEq0} and \eqref{LiouvilleEq} imply} immediately that $H$ is a first integral 
of the flow, that is, energy is conserved. In addition, any function commuting with $H$ is also a first integral.


\subsection{Canonical transformations and Liouville's theorem}\label{sectionLsympl}

Canonical transformations are an extremely important tool in classical mechanics, 
as they are strongly related to the symmetries and to the conserved quantities of the system and hence they are useful to simplify the equations of motion.
They can be classified in \emph{time-independent transformations}, the ones that preserve the form of the Hamiltonian function,
and \emph{time-dependent transformations}, which include time in the transformation and therefore are properly defined in an extended phase space. 
Here we consider time-independent transformations only. Time-dependent transformations are introduced below.

Canonical transformations are change of coordinates in the phase space that leave Hamilton's equations \eqref{Hameq2} invariant. 
From \eqref{Hameq1}, 
this amounts at finding a change of coordinates in the phase space that preserves the symplectic form $\Omega$~\cite{Arnold}.
This definition immediately yields a way to check whether a coordinate transformation is canonical. Given the transformation
$\left(q^{a},p_{a}\right)\to\left({Q^{a}},P_{a}\right)$, invariance of $\Omega$ implies the following conditions
	\begin{equation}
	\left\{ {Q^{a}} , {Q^{b}} \right\}_{(q^i,p_i)}=0\, , \quad 
	\left\{ P_a, P_b \right\}_{(q^i,p_i)}=0\, , \quad 
	\left\{ {Q^{a}} , P_{b} \right\}_{(q^i,p_i)}	=\delta^a_b\,.
	\end{equation}
As a consequence, canonical transformations also leave
the canonical form $\alpha$ 
invariant up to an exact differential, that is
	\beq\label{canonicalcondition}
	p_{a}\,\d q^{a}=P_{a}\,\d {Q^{a}}+\d F_{1} \, ,
	\eeq
where $F_{1}(q^a,{Q^{a}})$ is called the \emph{generating function} of the canonical transformation and obeys the relations  
	\begin{equation}\label{conditions-F}
	p_a = \frac{\partial F_{1}}{\partial q^a}, \quad P_a = - \frac{\partial F_{1}}{\partial {Q^{a}}}.
	\end{equation}
Furthermore, as $\Omega^{n}$ is the volume element of the phase space,
 then it follows that canonical transformations preserve the phase space volume.
A particular case of canonical transformations is the Hamiltonian evolution \eqref{Hameq2}.
In fact, symplectic Hamiltonian vector fields $X_{H}$ are the infinitesimal generators of canonical transformations.
Therefore Liouville's theorem
	 \beq\label{LiouvilleTh}
	 \pounds_{X_{H}}\Omega^{n}=0\,
	 \eeq
follows directly, where  $\pounds_{X_{H}}$ is the Lie derivative along the Hamiltonian vector field $X_{H}$~\cite{Arnold}.

\subsection{Time-dependent Hamiltonian systems}
\label{section-2.1}

For mechanical systems whose Hamiltonian depends explicitly on time the equations~\eqref{Hameq1} 
are no longer valid, since the differential of the Hamiltonian depends on time. 
 {Moreover, also in the case of time-independent systems, it is useful to consider time-dependent canonical transformations, 
for which the differential of the corresponding generating functions 
does not satisfy the canonical condition \eqref{canonicalcondition}.
In order to deal with time-dependent systems and time-dependent canonical transformations}, one usually extends the phase space with an extra dimension representing time.
The extended phase space $\Gamma^{\tiny\mbox{E}}=\Gamma\times\mathbb R$ is therefore a $(2n+1)$-dimensional manifold endowed with
a $1$-form
	\beq\label{PC1form}
	\eta_{\tiny\mbox{PC}}=p_{a}\d q^{a}-H\d t\,,
	\eeq
called the \emph{Poincar\'e-Cartan $1$-form},  {where the Hamiltonian $H$ can either depend explicitly on time or not\footnote{{Notice that
$(\Gamma^{\tiny\mbox{E}},\eta_{\tiny\mbox{PC}})$ is a contact manifold (cf. section \ref{subsec3.1}), 
but it is not the standard (natural) contactification of $(\Gamma,\Omega)$ (see \cite{arnold2001dynamical}), 
since
$\eta_{\tiny\mbox{PC}}$ depends on $H$ and hence on the system.}}.}
Then one proceeds to define a dynamics on $\Gamma^{\tiny\mbox{E}}$ that correctly extends Hamiltonian dynamics to the case where the Hamiltonian
depends explicitly on time. 
A direct calculation shows that the condition
	\beq\label{PCcondition}
	\d \eta_{\tiny\mbox{PC}}(X^{\tiny\mbox{E}}_{ H})=0
	\eeq
is satisfied if and only if the vector field $X^{\tiny\mbox{E}}_{H}$ in these coordinates takes the form 
	\beq\label{XHtilde}
	 X^{\tiny\mbox{E}}_{H}=X_{H}+\frac{\partial}{\partial t}\,,
	\eeq
where $X_{H}$ is given by \eqref{XH}.
Therefore the equations of motion for this vector field read
	\beq\label{Hameq3}
	\dot q^{a}=\frac{\partial H}{\partial p_{a}}\, , \qquad \dot p_{a}=-\frac{\partial H}{\partial q^{a}}\,, \qquad \dot t=1\,,
	\eeq
which are just Hamilton's equations \eqref{Hameq2}, augmented with the trivial equation $\dot{t}=1$. 
This makes clear that Hamilton's equation in the extended phase space~\eqref{Hameq3} are equivalent to the condition \eqref{PCcondition}.
It follows that the evolution of an arbitrary function $G\in C^{\infty}(\Gamma^{\tiny \mbox{E}})$ is given by
	\beq\label{Ev-TD}
	\frac{\d G}{\d t}=\{G,H\}_{(q^{a},p_{a})} + \frac{\partial G}{\partial t}\,,
	\eeq
 {and consequently for time-dependent Hamiltonian systems the} Hamiltonian itself is not conserved. 

Now let us study time-dependent canonical transformations and their generating functions.
To do so, we need to find a change of coordinates from $(q^{a},p_{a},t)$ to $(Q^{a},P_{a},t)$ 
that leaves the form of the extended Hamilton's equations~\eqref{Hameq3} unchanged. 
Since in condition \eqref{PCcondition} only the differential 
of $\eta_{\tiny\mbox{PC}}$ is involved, we find out that we can make a transformation that changes $\eta_{\tiny\mbox{PC}}$ 
by the addition of an exact differential, so that equation \eqref{PCcondition} is not affected.
Let us consider such transformation and write 
	\beq\label{PCdifference}
	p_{a}\d q^{a}-H\d t-(P_{a}\d Q^{a}-K\d t)=\d F_{1}\,,
	\eeq
where $K$ is a function on ${\Gamma^{\tiny\mbox{E}}}$ which is going to be the new Hamiltonian function after the transformation.
Let us further assume that we can choose coordinates in which $Q^{a}$ and $q^{a}$ are independent, so that the independent variables in \eqref{PCdifference} are
$(q^{a},Q^{a},t)$. We rewrite \eqref{PCdifference} as 
	\beq\label{PCdifference2}
	\left(p_{a}-\frac{\partial F_{1}}{\partial q^{a}}\right)\d q^{a}-\left(P_{a}+\frac{\partial F_{1}}{\partial Q^{a}}\right)\d Q^{a}+\left(K-H-\frac{\partial F_{1}}{\partial t}\right)\d t=0\,,
	\eeq
which implies that the \emph{generating function} of the canonical transformation
 $F_{1}(q^a,Q^a,t)$  satisfies the relations  
	\begin{equation}\label{conditions-Ftime}
	p_a = \frac{\partial F_{1}}{\partial q^a}, \quad P_a = - \frac{\partial F_{1}}{\partial Q^a}, \quad K=H+\frac{\partial F_{1}}{\partial t}\,.
	\end{equation}
Hamilton's equations~\eqref{Hameq3} in the new coordinates can be written as 
	\beq\label{Hameq4}
	\dot Q^{a}=\frac{\partial K}{\partial P_{a}}\, , \qquad \dot P_{a}=-\frac{\partial K}{\partial Q^{a}}\,, \qquad \dot t=1\,,
	\eeq
with $K$ the new Hamiltonian.

Systems with explicit time dependence are used for the effective description of dissipative systems 
within the Hamiltonian formalism. The idea is to  introduce a convenient time dependence into the Hamiltonian so that it reproduces the phenomenological equations
of motion with energy dissipation. 
As an example, let us consider the approach by Caldirola~\cite{caldirola} and Kanai~\cite{kanai} for a $1$-dimensional dissipative system with a friction force linear in the velocity. 
This model considers the time-dependent Hamiltonian
	\beq\label{CK-H}
	H_{\tiny\mbox{CK}}=e^{-\gamma t}\,\frac{p_{\tiny\mbox{CK}}^2}{2m}+e^{\gamma t} \, V(q_{\tiny\mbox{CK}})\, ,
	\eeq   
where $p_{\tiny\mbox{CK}}$ and $q_{\tiny\mbox{CK}}$ are the canonical coordinates in phase space, which are related to the physical positions and momenta by
the non-canonical transformation
	\beq\label{relphysCK}
	p_{\tiny\mbox{CK}} = e^{\gamma t} p\,, \quad q_{\tiny\mbox{CK}}=q .
	\eeq
It is easy to show that Hamilton's equations~\eqref{Hameq3} for $H$ as in~\eqref{CK-H} give the correct equation of motion for the position including the friction force, 
i.e. the damped Newton equation  
	\beq\label{New-CK}
	\ddot{q}+\gamma\,\dot{q}+\frac{1}{m}\frac{\partial V (q) }{\partial q}= 0\, .
	\eeq
 {However, although this model reproduces the correct phenomenological equation of motion, it has the drawback that in order to describe
dissipative systems 
one needs to take into account the non-canonical relationship~\eqref{relphysCK} between canonical and physical 
quantities.
As a consequence, at the quantum level this model has generated quite a dispute on whether it can describe a dissipative system without violating the Heisenberg uncertainty principle; 
we refer
to e.g. the discussion in~\cite{greenberger1979critique,schuch1997nonunitary, um2002quantum, schuch2011dynamical, cruz2015time, cruz2016time} and references therein.}

%
%


\subsection{Hamilton-Jacobi formulation}
\label{sec2.4}

The Hamilton-Jacobi formulation is a powerful tool which enables to re-express Hamilton's equations 
in terms of a single partial differential equation whose solution, a function of the configuration space, has all the necessary information to obtain 
the trajectories of the mechanical system. 
Moreover, this formulation gives rise to a new and more geometric point of view that allows to relate classical mechanics with wave phenomena
and thus with quantum mechanics. 

The Hamilton-Jacobi equation can be introduced as a special case of a time-dependent canonical transformation \eqref{conditions-Ftime}. 
Consider the case in which the new Hamiltonian $K$ vanishes
and write the generating function $F_{1}$ in this particular case as $S$. Using \eqref{conditions-Ftime}, we can write the \emph{Hamilton-Jacobi equation}
	\beq\label{HJ1}
	H\left(q^{a},\frac{\partial S}{\partial q^{a}},t\right)=-\frac{\partial S}{\partial t}\,.
	\eeq
A complete solution $S(q^a,t)$, called \emph{Hamilton's principal function},  
determines completely the dynamics of the system~\cite{goldstein2002classical}. 
Besides, since $K=0$ for such transformation, it is clear from \eqref{Hameq4} that Hamilton's equations in the new coordinates read
	\beq\label{Hameq6}
	\dot Q^{a}=0, \qquad \dot P_{a}=0\,.
	\eeq
Therefore, the new system of coordinates moves along the Hamiltonian flow. In fact, the functions $Q^{a}(q^{i},p_{i},t)$ and $P_{a}(q^{i},p_{i},t)$ are 
(generalized) Noether invariants associated
with the Noether symmetries ${\partial}/{\partial P_{a}}$ and ${\partial}/{\partial Q^{a}}$ respectively \cite{aldaya2014contact}.
Finally, the time derivative of Hamilton's principal function is given by
	\beq\label{dSdt}
	\dot S=\frac{\partial S}{\partial q^{a}}\dot q^{a}+\frac{\partial S}{\partial t}=p_{a}\dot q^{a}-H\,
	\eeq
where in the second identity we have used both \eqref{conditions-Ftime} and \eqref{HJ1}.
Since the right hand side of \eqref{dSdt} is the Lagrangian of the system, one concludes that
	\begin{equation}\label{actionf}
 	S(q^a,t) = \int L(q^a,\dot{q}^a,t) \d t \,,
	\end{equation}
i.e.~that Hamilton's principal function is the action, up to an undetermined additive constant~\cite{goldstein2002classical}.


\section{Contact mechanics of dissipative systems}
\label{section-3}
So far we have only reviewed the standard Hamiltonian description of mechanical systems.
In this section we introduce the formalism of contact Hamiltonian mechanics and show that it can be applied to describe both non-dissipative and dissipative systems.
 {Some of the material in sections \ref{subsec3.1} and \ref{subsec3.3} has been already presented in \cite{bravetti2015liouville,PhysRevE.93.022139}.}

\subsection{Time-independent contact Hamiltonian mechanics}
\label{subsec3.1}
A \emph{contact manifold} $\mathcal T$ is a {$(2n+1)$-dimensional} manifold endowed with a 1-form $\eta$, called the \emph{contact form}, 
that satisfies the condition~\cite{arnold2001dynamical}
\begin{equation}\label{integraxx}
   \eta \wedge (\d \eta)^n \neq 0\,.
\end{equation}
The left hand side in \eqref{integraxx} provides the \emph{standard volume form} on $\mathcal T$, analogously to $\Omega^{n}$ for the symplectic case.
Hereafter we assume that the phase space  {of time-independent} 
mechanical systems (both dissipative and non-dissipative) is a contact manifold,  {called the 
\emph{contact phase space}\footnote{{The reader familiar with the geometric representation of quantum mechanics might 
notice the similarity between the concepts of contact phase space and quantum phase space. 
Both of them may be seen as a fiber bundle over the symplectic phase space~\cite{brody2001geometric,isidro2002duality,venuti2007quantum,heydari2015geometry}.}}}, 
and that the equations of motion are always 
given by the so-called
contact Hamiltonian equations. We show that in this way one can construct a Hamiltonian formalism for
any mechanical system.
First let us define the dynamics in the phase space $\mathcal T$. 
 {Given the $1$-form $\eta$, one can associate to every
differentiable function {$\mathscr{H}:\mathcal{T}\rightarrow\mathbb{R}$}, 
 a vector field $X_{\mathscr{H}}$, called the \emph{contact Hamiltonian vector field generated by $\mathscr{H}$},
defined through the two (intrinsic) relations
	\beq\label{isomorphismLiealg}
	\pounds_{X_{\H}}\eta=f_{\H}\,\eta \qquad \text{and} \qquad -\H = \eta\left(X_\mathscr{H}\right)\,,
	\eeq
where $f_{\H}\in C^{\infty}(\mathcal T)$ is a function depending on $\H$ to be fixed below, cf.~equations~\eqref{reebH} and \eqref{dHdS}, 
and $\mathscr{H}$ is called the \emph{contact Hamiltonian}~\cite{Mrugaa:2000aa,arnold2001dynamical,Boyer}.
The first condition in \eqref{isomorphismLiealg} means that $X_{\H}$ generates a contact transformation (see section \ref{subsec3.3} below), while the second condition
guarantees that it is generated by a Hamiltonian function.
Using Cartan's identity~\cite{Arnold} 
	\begin{equation}
	\label{cartan}
	\pounds_{X_{\mathscr{H}}}\eta=\d \eta(X_{\mathscr{H}})+\d[\eta(X_{\mathscr{H}})]
	\end{equation}
 and the second condition in~\eqref{isomorphismLiealg}, it follows that
	\beq\label{eqdH}
	\d\H=\d \eta(X_{\H})-\pounds_{X_{\H}}\eta\,,
	\eeq
from} which it is clear that the definition of a contact Hamiltonian vector field generalizes that of a symplectic Hamiltonian vector field to the case where
the defining $1$-form is not preserved along the flow [cf.~equations \eqref{eqdH} and \eqref{Hameq1}].

An example of a contact manifold is the extended phase space ${\Gamma^{\tiny\mbox{E}}}$
that we have introduced in section \ref{section-2.1} in order to account for time-dependent Hamiltonian systems. 
In fact, it is easy to prove that the Poincar\'e-Cartan $1$-form $\eta_{\tiny\mbox{PC}}$ satisfies the condition \eqref{integraxx} and therefore it defines a
contact structure on~$\Gamma^{\tiny\mbox{E}}\,$.

Associated with the definition of the contact $1$-form on a contact manifold, there is another fundamental object called the \emph{Reeb vector field} $\xi$, which is defined
intrinsically by the conditions
	\beq\label{reeb}
	\eta(\xi)=1, \qquad \d\eta(\xi)=0\,.
	\eeq
It can be shown that such vector field is unique and that it defines at every point a `vertical' direction with respect to the horizontal distribution $\mathcal{D}={\rm ker}(\eta)$. 
 {Finally, using \eqref{eqdH} and \eqref{reeb}, it is easy to prove that the two conditions in \eqref{isomorphismLiealg} imply
	\beq\label{reebH}
	f_{\H}=-\xi(\H)\,.
	\eeq
}

It is always possible to find a set of local (Darboux) coordinates $(q^{a},p_{a},S)$ for $\mathcal{T}$~\cite{Arnold}, to which we refer as \emph{contact coordinates}, 
such that the 1-form $\eta$ and the Reeb vector field $\xi$ can be written as
	\begin{equation}\label{1stform}
	\eta = \d S - p_a \d q^a, \qquad \xi=\frac{\partial}{\partial S}\,.
	\end{equation}
 {We remark that $\eta$ as in \eqref{1stform} is the standard (natural) contactification
of a symplectic manifold whose symplectic structure is exact, as defined e.g.~in~\cite{arnold2001dynamical} and that
the second expression in \eqref{1stform} directly implies that in these coordinates 
	\beq\label{dHdS}
	f_{\H}=-\xi(\H)=-\frac{\partial \H}{\partial S}\,.
	\eeq
Besides,}
in these coordinates, the contact Hamiltonian vector field $X_{\mathscr{H}}$ takes the form 	
	\beq
	\begin{split}
	\label{generic.ham}
	X_\mathscr{H} =& \left(p_a \frac{\partial \mathscr{H}}{\partial p_a}-\H  \right)\frac{\partial}{\partial S} 
		- \left(p_a \frac{\partial \mathscr{H}}{\partial S}+\frac{\partial \mathscr{H}}{\partial q^a} \right)\frac{\partial }{\partial p_a} 
		+ \left(\frac{\partial \mathscr{H} }{\partial p_a} \right)\frac{\partial }{\partial q^a}\,.
	\end{split}
	\eeq	
According to equation \eqref{generic.ham}, the flow of $X_{\mathscr{H}}$ can be explicitly written in contact coordinates as
	\begin{empheq}{align}
	\label{z3}
	& \dot{q}^{a} \,= \frac{\partial \mathscr{H} }{\partial p_a}\,,\\
	\label{z2}
	& \dot{p}_{a} \,=  -\frac{\partial \mathscr{H}}{\partial q^a} - p_a \frac{\partial \mathscr{H}}{\partial S}\,,\\
	 \label{z1}
	& \dot S \,\,= p_a \frac{\partial \mathscr{H}}{\partial p_a} -\H\,.
	\end{empheq}
	
The similarity of equations \eqref{z3}-\eqref{z1} with Hamilton's equations of symplectic mechanics~\eqref{Hameq2} is manifest. 
In fact, these are the generalization of Hamilton's equations to a contact manifold. In particular, when $\mathscr{H}$ does not depend on $S$, 
 equations \eqref{z3} and \eqref{z2} give exactly Hamilton's equations in the symplectic phase space and $\H$ is an integral of motion.
Finally, the remaining equation \eqref{z1} in this case is the usual definition of Hamilton's principal function -- cf.~equation~\eqref{dSdt}. 
Therefore  \eqref{z3}-\eqref{z1} generalize the equations of motion for the positions, the momenta and Hamilton's principal function of the standard 
Hamilton's theory and can include a much larger class of models, such as the dynamics of basic dissipative systems (that we consider below) 
and that of systems in equilibrium with a heat bath, i.e.~the so-called ``thermostatted dynamics"~\cite{PhysRevE.93.022139,TuckermanBook,evansbook,tapias2016geometric}
(see also~\cite{Bravetti2015377,shinitiro1,shinitiro2} for applications to non-equilibrium thermodynamics).
	
As an example, given the ($1$-dimensional) contact Hamiltonian system
	\beq\label{H-lin-S}
	\H _S = \frac{p^2}{2m} + V(q) + \gamma\,S
	\eeq
where $V(q)$ is the mechanical potential and $\gamma$ is a constant parameter, the equations of motion~\eqref{z3}-\eqref{z1} read 
	\begin{eqnarray}
	 \dot{q} &=& \frac{p}{m}\, ,\label{linearq} \\
	 \dot{p} &=& -\frac{\partial V(q)}{\partial q} - \gamma \, p\, , \label{linearp}\\
	 \dot {S}&=& \frac{p^2}{2 m}-V(q)-\gamma\, S \label{linearS}\, .
	\end{eqnarray}
From~\eqref{linearq} and \eqref{linearp} it is easy to derive the damped Newtonian equation~\eqref{New-CK},
which describes all systems with a friction force that depends linearly on the velocity.
Notice that the derivation through the use of contact Hamiltonian dynamics guarantees that the canonical and physical momenta and positions coincide, contrary to what happens 
in the case of a description by means of explicit time dependence, as for instance in the Caldirola-Kanai model~\eqref{CK-H}.

Before concluding this section, let us remark an important difference between our approach and previous uses of contact geometry to describe non-conservative systems.
 As we showed in section~\ref{section-2.1}, the evolution of a 
 non-conservative mechanical system whose Hamiltonian depends explicitly on time is usually given in the extended phase space $\Gamma^{\tiny\mbox{E}}$,
 endowed with the Poincar\'e-Cartan  $1$-form~\eqref{PC1form}, which provides the contact $1$-form
 for $\Gamma^{\tiny\mbox{E}}$~\cite{abraham1978foundations}. 
 Usual treatments of time-dependent mechanical systems give the dynamics as in~\eqref{XHtilde}. 
Therefore, according to \eqref{isomorphismLiealg}
one finds that the corresponding contact Hamiltonian is 
	\beq\label{contactHtimedep}
	-\H=\eta_{\tiny\mbox{PC}}(X_{H}^{\tiny\mbox{E}})=p_{a}\frac{\partial H}{\partial p_{a}}-H=[H]\,,
	\eeq
where $[H]$ stands for the total Legendre transform of $H$~\cite{Arnold}. 
Moreover, from the condition \eqref{PCcondition}, defining the Hamiltonian dynamics in $\Gamma^{\tiny\mbox{E}}$, and
from the definition of the Reeb vector field \eqref{reeb}, one finds immediately that $X_{H}^{\tiny\mbox{E}}$ is proportional to the Reeb vector field
 $\xi$ in the extended phase space, 
the proportionality being given by $-\H$. 
One concludes then that any time-dependent mechanical system can be described in $\Gamma^{\tiny\mbox{E}}$ 
by the contact Hamiltonian vector
field
	\beq\label{Xcontacttimedep}
	X_{\H}=-\H\xi=\left(p_{a}\frac{\partial H}{\partial p_{a}}-H\right)\frac{\partial}{\partial S}\,.
	\eeq
The flow of this vector field in contact coordinates $(Q^{a},P_{a},S)$ is
	\begin{eqnarray}
	&\,\dot Q^{a}=&\,0\\
	&\dot P_{a}\,=&\,0\\
	&\dot{S}\,\,\,=&[H]\,,
	\end{eqnarray}
which coincides with the flow \eqref{Hameq6}-\eqref{dSdt}, 
i.e.~the natural evolution in the adapted coordinates found after performing the proper (Hamilton-Jacobi) time-dependent canonical transformation~\cite{aldaya2014contact}. 

In this work we decide not to take this description for time-dependent Hamiltonian systems.
In fact, we always consider here time-independent symplectic systems as embedded into the contact phase space $\mathcal{T}$
and we use the mechanical Hamiltonian $H_{\tiny\mbox{mec}}(q^{a},p_{a})$
as a contact Hamiltonian to write the equations of motion in the form~\eqref{z3}-\eqref{z1}.
It is easy to see that since $H_{\tiny\mbox{mec}}$ does not depend on the additional variable $S$ explicitly, the equations of motion thus derived are Hamilton's equations~\eqref{Hameq2}
for the time-independent case.
In order to consider the time-dependent case, we develop in section~\ref{Tdepcontact} a formalism for time-dependent contact Hamiltonian systems
and then again we recover standard mechanical systems given by a mechanical Hamiltonian of the type $H_{\tiny\mbox{mec}}(q^{a},p_{a},t)$
as a particular case of the more general time-dependent contact Hamiltonian evolution, thus obtaining again the correct equations~\eqref{Hameq3} as a particular case.
The two main advantages of our perspective are that we can always identify the canonical variables $(q^{a},p_{a})$ with the physical ones
and that -- as we show below -- we can classify mechanical systems as \emph{dissipative} or \emph{non-dissipative} in terms of
the contraction of the phase space volume.


\subsection{Time evolution of the contact Hamiltonian and mechanical energy}\label{subsec3.2}

In this section we derive the evolution of the contact {Hamiltonian} and {the} mechanical energy for a system 
{evolving according to the} contact Hamiltonian equations~\eqref{z3}-\eqref{z1}
and we show that there is a constant of motion that can help to simplify the solution of the dynamics in particular cases.

Given any function in the contact phase space $\mathscr{F}\in C^{\infty}(\mathcal T)$, its evolution according to equations~\eqref{z3}-\eqref{z1} is given by 
	\begin{eqnarray}\label{ContactEvolution}
	\frac{\d \mathscr{F}}{\d t}&=&X_\mathscr{H}[ \mathscr{F} ]\nonumber \\
	&=& -\mathscr{H}\,\frac{\partial \mathscr{F}}{\partial S}+p_a\left[ \frac{\partial \mathscr{F}}{\partial S}\frac{\partial \mathscr{H}}{\partial p_a}
	-\frac{\partial \mathscr{F}}{\partial p_a}\frac{\partial \mathscr{H}}{\partial S}\right] +
	\frac{\partial \mathscr{F}}{\partial q^a}\frac{\partial \mathscr{H}}{\partial p_a}-\frac{\partial \mathscr{F}}{\partial p_a}\frac{\partial \mathscr{H}}{\partial q^a}\nonumber\\
	&=&-\mathscr{H}\,\frac{\partial \mathscr{F}}{\partial S}+p_a\left\{\mathscr{F},  \mathscr{H}\right\}_{(S,p_a)}+ \left\{\mathscr{F},  \mathscr{H}\right\}_{(q^a,p_a)}\,,
	\end{eqnarray}
where $\{ \,\, , \, \}_{(q^{a},p_{a})}$ is the standard Poisson bracket as in \eqref{LiouvilleEq} and the remaining terms are contact corrections. 
 {We point out that  the bracket $\{ \,\, , \, \}_{(S,p_{a})}$ is just a shorthand notation
and we do not provide any intrinsic definition for it.}
We say that a function $\mathscr{F}\in C^{\infty}(\mathcal T)$ is a \emph{first integral}  {(or \emph{invariant})} of the contact dynamics given by $X_{\mathscr{H}}$ if
$\mathscr{F}$ is constant along the flow of $X_{\mathscr{H}}$, that is if $X_{\mathscr{H}}[\mathscr{F}]=0$.
From the above equations, it follows that the evolution of the contact Hamiltonian function along its flow is
	\beq\label{flowofh}
	\frac{\d \mathscr{H}}{\d t}=-\mathscr{H}\,\frac{\partial \mathscr{H}}{\partial S}\,.
	\eeq
Therefore in general $\H$ is constant if and only if $\H=0$
or if $\mathscr{H}$ does not depend on $S$. The latter case corresponds to a non-dissipative mechanical system, for which
 $\H=H_{\tiny\mbox{mec}}(q^{a},p_{a})$ and thus the mechanical energy is conserved.
Let us consider a more general case, in which
	\beq\label{H-diss}
	\H = H_{\tiny \mbox{mec}}(q^{a},p_{a})+ h(S)\,, 
	\eeq 
where $H_{\tiny \mbox{mec}}(q^{a},p_{a})$ is the mechanical energy of the system and $h(S)$ characterizes effectively the interaction with the environment.
From~\eqref{ContactEvolution}, the evolution of the mechanical energy is 
	\beq\label{dis-en}
	\frac{\d H_{\tiny \mbox{mec}}}{\d t} =-p_{a}\frac{\partial H_{\tiny\mbox{mec}}}{\partial p_{a}}\,h'(S)\,, 
	\eeq
from which it is clear that $h$ is a potential that generates dissipative forces.
For example, in the case of mechanical systems with linear friction represented by the contact Hamiltonian~\eqref{H-lin-S}, we see that the rate of dissipation of the mechanical energy is
	\beq\label{dissElinear}
	\frac{\d H_{\tiny \mbox{mec}}}{\d t} = - m \, \gamma \, \dot{q}^2 \, , 
	\eeq
which agrees with standard results {based on Rayleigh's dissipation function}~\cite{goldstein2002classical}. 
Furthermore, the evolution of the contact Hamiltonian \eqref{H-diss} can be formally obtained from~\eqref{flowofh} to be
	\beq\label{TE-H}
	\H(t) = \H_0 \exp\left\{ - \int^t_{0} h'(S)\, \d \tau \right\}\,.
	\eeq
In the example of the mechanical system with linear friction, the contact Hamiltonian~\eqref{H-lin-S}
depends linearly on $S$ and therefore its evolution reads
	\beq\label{ener-tot}
	\H_{S}(t)=\H_{S,0} \, e^{ - \gamma t}\, ,
	\eeq
{where $ \H_{S,0}$ is the value of $\H_{S}$ at $t=0$.}

Equation~\eqref{TE-H} introduces the constant of motion $\H_0$, which eliminates one degree of freedom from the equations of motion~\eqref{z3}-\eqref{z1}. 
In fact, inserting the contact Hamiltonian \eqref{H-diss} into \eqref{TE-H} one obtains in general
	\beq \label{41}
	H_{\tiny \mbox{mec}}(q^a,p_a)+ h(S) =  \H_0 \exp\left\{ - \int^t_{0} h'(S) \, \d \tau \right\}\, ,
	\eeq
and in principle one can solve this equation for any contact coordinate. 
In particular, it is possible to solve~\eqref{41} to obtain $S$ as a function of $q^{a},p_{a}$ and $t$
and therefore the {solution} of the system~\eqref{z3}-\eqref{z1} 
then amounts only to solve the $2n$ equations for the momenta and the positions, as in the standard symplectic case.
For example, with $\H_{S}$ as in~\eqref{H-lin-S} one obtains
	\beq\label{Slinearcase}
	S(q,p,t) = \frac{1}{\gamma}\left[ \H_{S,0} \, e^{ - \gamma t} - \frac{p^2}{2m} - V(q) \right] \,.
	\eeq


\subsection{Contact transformations and Liouville's theorem}\label{subsec3.3}

In the preceding sections we have introduced the contact phase space for  {time-independent}
mechanical systems, equipped with the local coordinates $(q^{a},p_{a},S)$, called contact coordinates. 
In these variables the equations of motion are expressed in terms of the contact Hamiltonian equations  \eqref{z3}-\eqref{z1} and the contact form is expressed as in~\eqref{1stform}. 
As in the symplectic case, we are now interested in introducing those transformations that leave the contact structure unchanged, which are known as \emph{contact transformations}~\cite{Arnold,Boyer}. 
Here we consider only time-independent contact transformations and {in the next subsection} we introduce the time-dependent case.

A \emph{contact transformation} is {a} transformation that leaves the contact form invariant up to multiplication by a conformal factor~\cite{rajeev2008quantization,rajeev2008hamilton}, that is
	\beq\label{contacttrdef}
	\tilde\eta=f\,\eta\,.
	\eeq
From~\eqref{contacttrdef}, an arbitrary transformation of coordinates from $(q^a,p_a,S)$ to $(\tilde{Q}^a,\tilde{P}_a,\tilde S)$ is a contact transformation if
	\begin{equation}\label{contact-transformation}
	f (\d S-p_a \d q^a)=\d{\tilde S}-\tilde{P}_a\d \tilde{Q}^a\,,
	\end{equation}
which is equivalent to
	\begin{eqnarray}\label{contact-conditions}
	f&=&\frac{\partial \tilde{S}}{\partial S}-\tilde{P}_a\frac{\partial \tilde{Q}^a}{\partial S}\label{contact1}\\
	-f p_i&=&\frac{\partial \tilde{S}}{\partial q^i}-\tilde{P}_a\frac{\partial \tilde{{Q}}^a}{\partial q^i}\label{contact2}\\
	0&=&\frac{\partial \tilde{S}}{\partial p_i}-\tilde{P}_a\frac{\partial \tilde{Q}^a}{\partial p_i}\label{contact3}\,.
	\end{eqnarray} 

As in the standard symplectic theory, we can obtain the generating function of a contact transformation. 
Assuming that the coordinates $(q^a, \tilde{Q}^a,S)$ are independent, we compute the differential 
of the generating function $\tilde{S}(q^a,\tilde{Q}^a,S)$, namely
	\beq\label{dStilde}
	\d \tilde{S}=\frac{\partial \tilde{S}}{\partial S}\d S + \frac{\partial \tilde{S}}{\partial q^a}\d q^a + \frac{\partial \tilde{S}}		{\partial \tilde{Q}^a}\d \tilde{Q}^a\,.
	\eeq
Substituting~\eqref{dStilde} into~\eqref{contact-transformation} we obtain the following conditions for $\tilde S$
	\beq\label{gen-fun}
	f=\frac{\partial \tilde{S}}{\partial S}\, , \quad f \, p_a = - \frac{\partial \tilde{S}}{\partial q^a}\, , \quad \tilde P_a = \frac{\partial 			\tilde{S}}{\partial \tilde{Q}^a}\,.
	\eeq 
In particular, for contact transformations with $f =1$ the conditions in~\eqref{gen-fun} imply that the generating function has the form 
	\beq
	\tilde S = S - F_1(q^a, \tilde{Q}^a)\,, 
	\eeq	
where $F_{1}(q^{a},\tilde{Q}^{a})$ is the generating function of a symplectic canonical transformation, cf.~equation \eqref{conditions-F}. 
This result is remarkable, since it implies that all canonical transformations are a special case of contact transformations corresponding to $f=1$.

While canonical transformations preserve the symplectic volume form $\Omega^{n}$, we show now that contact transformations induce a re-scaling of the
contact volume form $\eta\wedge\left(\d\eta\right)^{n}$.
Let us assume that we have a transformation that induces the change $\tilde\eta=f\, \eta$; then $\d\tilde\eta=\d f\wedge \eta+f\,\d\eta$. It follows that 
	\beq\label{rescalingV}
	\tilde\eta\wedge\left(\d\tilde\eta\right)^{n}=f^{n+1}\eta\wedge\left(\d\eta\right)^{n}\,,
	\eeq
i.e.~the volume form is rescaled by a term $f^{n+1}$, with $f$ given in general in~\eqref{contact1}. Note that canonical transformations are a special case with $f=1$
and therefore they preserve the contact volume form.

Finally, applying the contact Hamiltonian vector field $X_{\H}$ to $\eta$,  {we see from \eqref{isomorphismLiealg} and \eqref{dHdS} that 
	\beq\label{Lieeta}
	\pounds_{X_{\H}}\eta=f_{\H}\eta=-\frac{\partial \H}{\partial S}\eta\,.
	\eeq
}
 Comparing \eqref{Lieeta} with \eqref{contacttrdef},
  {we conclude} that contact Hamiltonian vector fields are the infinitesimal generators of contact transformations  \cite{rajeev2008quantization,rajeev2008hamilton}. 
 Again, this is the analogue of the fact that symplectic Hamiltonian vector fields
 are the infinitesimal generators of canonical transformations.
 Moreover, equation \eqref{Lieeta} also implies that the volume element contracts (or expands) along the contact Hamiltonian flow according to \cite{bravetti2015liouville}
 	\beq\label{LieVolume}
	\pounds_{X_{\H}}\left(\eta\wedge(\d\eta)^{n}\right)=-(n+1)\frac{\partial \H}{\partial S}\left(\eta\wedge(\d\eta)^{n}\right)\,,
	\eeq
which means that the contact flow has a non-zero divergence 
	\beq\label{divergence}
	{\rm div}(X_{\H})=-(n+1)\frac{\partial \H}{\partial S}
	\eeq
and therefore Liouville's theorem~\eqref{LiouvilleTh} does not hold.
However,
an analogous statement of Liouville's theorem for contact flows has been proved in \cite{bravetti2015liouville}.  
In fact, although the volume element $\eta\wedge\left(\d\eta\right)^{n}$ is not preserved
 along the contact Hamiltonian flow, nevertheless a unique invariant measure depending only on $\H$ can be found  whenever $\H\neq0$, 
 given by 
 	\beq\label{invmeasure1}
	\d\mu=|\H|^{-(n+1)}\left(\eta\wedge\left(\d\eta\right)^{n}\right)\,,
	\eeq 
where the absolute value $|\cdot|$ has been introduced in order to ensure that the probability distribution is positive.
As it provides an invariant measure for the flow, this is the analogue of Liouville's theorem for contact Hamiltonian flows.	

Since the presence of a non-zero divergence is usually interpreted as a sign of dissipation~\cite{daems1999entropy,gallavotti2004nonequilibrium},
here we classify systems as~\emph{non-dissipative}  or \emph{dissipative} depending on whether the divergence~\eqref{divergence} of the associated 
dynamics vanishes or not. 


\subsection{Time-dependent contact Hamiltonian systems}\label{Tdepcontact}

In the preceding sections we have seen that contact Hamiltonian mechanics can account for the dynamics of mechanical systems with dissipation
and we have proven some results that extend the symplectic formalism to the contact case. However, so far we have considered only time-independent systems.
Now we introduce contact Hamiltonian systems that explicitly depend on time. 
The results of this and the following sections are all new.

To begin, let us extend the contact phase space by adding the time variable to it. 
Therefore we have an extended manifold ${\mathcal T^{\tiny\mbox{E}}} = \mathcal T\times \mathbb R$ with natural coordinates
derived from contact coordinates as $(q^{a},p_{a},S,t)$. Then we extend the contact $1$-form \eqref{1stform} to the $1$-form 
	\beq\label{extended1-form}
	{\eta}^{\tiny\mbox{E}}=\d S-p_{a}\d q^{a}+\mathscr{H}\d t\,,
	\eeq
where $\H$ is the contact Hamiltonian, that in this case is allowed to depend on $t$ too.
Notice that whenever $\H$ depends on $S$, $\d \eta^{\tiny\mbox{E}}$ is non-degenerate (and closed) and therefore $({\mathcal T^{\tiny\mbox{E}}},\d \eta^{\tiny\mbox{E}})$
is a symplectification of $(\mathcal T,\eta)$. However, such symplectification is not the standard (natural) one defined e.g.~in~\cite{arnold2001dynamical}.
Our symplectification depends on the Hamiltonian of the system as it is clear from equation~\eqref{extended1-form}. This is the same as it happens with 
the contactification of the symplectic phase
space given by the Poincar\'e-Cartan $1$-form \eqref{PC1form}.
Besides, the coordinates $(q^{a},p_{a},S,t)$ are non-canonical coordinates for $\d \eta^{\tiny\mbox{E}}$,  as it is easy to check.
%

{Now we want to define the dynamics on $\mathcal T^{\tiny\mbox{E}}$. To do so, we set the two (intrinsic) simultaneous conditions
	\beq\label{t-dep-dynamics-conditions0}
	\pounds_{X^{\tiny\mbox{E}}_{\H}}\eta^{\tiny\mbox{E}}=g_{\H}\,\eta^{\tiny\mbox{E}} \qquad \text{and} \qquad  \eta^{\tiny\mbox{E}}\left(X^{\tiny\mbox{E}}_\mathscr{H}\right)=0\,,
	\eeq
with $g_{\H}\in C^{\infty}(\mathcal T^{\tiny\mbox{E}})$ a function depending on $\H$ to be fixed below, cf.~equation \eqref{conditions74}.
Notice that \eqref{t-dep-dynamics-conditions0} is the natural extension of \eqref{isomorphismLiealg} to $\mathcal T^{\tiny\mbox{E}}$.
We argue that these two conditions define a vector field ${X}^{\tiny\mbox{E}}_{\H}$ on ${\mathcal T}^{\tiny\mbox{E}}$ which is completely equivalent to the contact Hamiltonian flow \eqref{generic.ham}.
To prove this, let us first
use Cartan's identity \eqref{cartan} to re-write~\eqref{t-dep-dynamics-conditions0} as
	\beq\label{t-dep-dynamics-conditions}
	\d{\eta^{\tiny\mbox{E}}}({X}^{\tiny\mbox{E}}_{\H})=g_{\H}\eta^{\tiny\mbox{E}} \qquad {\rm and} \qquad {\eta^{\tiny\mbox{E}}}({X}^{\tiny\mbox{E}}_{\H})=0\,.
	\eeq
Then we use the second}
  condition in \eqref{t-dep-dynamics-conditions}.
In local coordinates we can write this condition as
	\beq\label{cond1}
	\left(\d S-p_{a}\d q^{a}+\mathscr{H}\d t\right)\left({X}^{S}\frac{\partial}{\partial S}+{X}^{q^{a}}\frac{\partial}{\partial q^{a}}+{X}^{p_{a}}\frac{\partial}{\partial p_{a}}+{X}^{t}\frac{\partial}{\partial t}\right)=0\,,
	\eeq
where the ${X}^{i}$ are the general components of the vector field $X_{\H}^{\tiny\mbox{E}}$ in these coordinates. We are free to fix a normalization for ${X}_{\H}^{\tiny\mbox{E}}$ such that
${X}^{t}=1$. Now condition \eqref{cond1} yields 
	\beq\label{cond1bis}
	{X}^{S}=p_{a}{X}^{q^{a}}-\H\,.
	\eeq
Using \eqref{cond1bis} we can write  the {first} condition in~\eqref{t-dep-dynamics-conditions} as
 {
	\beq
	\d\eta^{\tiny \mbox{E}}\left([p_{a}\,X^{q^{a}} - \H]\frac{\partial}{\partial S}
	+X^{q^{a}}\frac{\partial}{\partial q^{a}}+X^{p_{a}}\frac{\partial}{\partial p_{a}}+\frac{\partial}{\partial t}\right)
	=g_{\H}\eta^{\tiny\mbox{E}}\, ,
	\eeq
and, after a direct calculation, one arrives at
	\beq\label{conditions74}
	g_{\H}= -\frac{\partial \H}{\partial S}\, , \quad
	X^{q^{a}} = \frac{\partial \H}{\partial p_{a}}\, , \quad
	X^{p_{a}} = - \frac{\partial \H}{\partial q^{a}}-p_{a}\frac{\partial \H}{\partial S}.
	\eeq
}
Finally, considering all the above conditions, we can write the resulting vector field $X_{\H}^{\tiny\mbox{E}}$ satisfying both conditions in \eqref{t-dep-dynamics-conditions} in its general form as
	\beq\label{Xtilde}
	{X}_{\H}^{\tiny\mbox{E}}=X_{\H}+\frac{\partial}{\partial t}\,,
	\eeq 
with $X_{\H}$ given by \eqref{generic.ham}. From this it is immediate to recognize that the equations of motion given by such field on ${\mathcal T}^{\tiny\mbox{E}}$ are the same as those of
the contact Hamiltonian vector field \eqref{generic.ham}, with the addition of the trivial equation $\dot{t}=1$.
We call a system defined by a contact Hamiltonian $\H(q^{a},p_{a}, S, t)$ and by the vector field $X_{\H}^{\tiny\mbox{E}}$ of the form~\eqref{Xtilde}
a \emph{time-dependent contact Hamiltonian system}\footnote{We emphasize that, although $(\mathcal T^{\tiny\mbox{E}},\d \eta^{\tiny\mbox{E}})$ is a symplectic manifold,
the flow $X_{\H}^{\tiny\mbox{E}}$ has a non-vanishing divergence and therefore it is not a standard symplectic Hamiltonian dynamics, nor it can 
be introduced in terms of the usual Dirac formalism for time-independent constrained systems.}.
From~\eqref{Xtilde} and~\eqref{ContactEvolution}
it follows that the evolution of any function $\mathscr{F}\in C^{\infty}(\mathcal T^{\tiny\mbox{E}})$ under the dynamics given by a time-dependent contact
Hamiltonian system reads
	\beq\label{evol-fun-TD}
	\frac{\d \mathscr{F}}{\d t} = -\mathscr{H}\,\frac{\partial \mathscr{F}}{\partial S}
	+p_a\left\{\mathscr{F},  \mathscr{H}\right\}_{(S,p_a)}
	+ \left\{\mathscr{F},  \mathscr{H}\right\}_{(q^a,p_a)}
	+\frac{\partial \mathscr{F}}{\partial t}\, .
	\eeq

Now that we have found a formal prescription to write the equations of motion for time-dependent contact Hamiltonian systems, let us
discuss time-dependent contact transformations and their generating functions.
\emph{Time-dependent contact transformations} are transformation of coordinates 
	\beq\label{generalTrTE}
	(q^a,p_a,S,t) \to (\tilde{Q}^a,\tilde{P}_a,\tilde S,t)\, ,
	\eeq
that leave the equations of motion, i.e.~the vector field $X_{\H}^{\tiny\mbox{E}}$, invariant.
 By definition, this amounts at finding a transformation that leaves both conditions in~\eqref{t-dep-dynamics-conditions} unchanged.
To find such a transformation, we start with the  {second} condition and write the invariance as the fact that the transformed extended $1$-form must have the same form as the original one
up to multiplication by a non-zero function $f$, that is
	\beq\label{canonicalcontact1}
	f\left(\d S-p_{a}\d q^{a}+\mathscr{H}\d t\right) = \d \tilde{S}-\tilde{P}_{a}\d {\tilde{Q}^{a}}+\mathscr{K}\d t\,,
	\eeq
where $\mathscr{K}$ is a function on ${\mathcal T}^{\tiny\mbox{E}}$ which is going to be the new contact Hamiltonian in the transformed coordinates.
This condition provides a way to check whether a transformation of the type~\eqref{generalTrTE}
is a time-dependent contact transformation. Indeed, 
inserting the differentials of $\tilde{Q}^a$ and $\tilde{S}$ into \eqref{canonicalcontact1} one obtains the standard conditions~\eqref{contact1}-\eqref{contact3} 
for a time-independent contact transformation, together with the following rule for the transformation of the Hamiltonians
	\beq\label{relSHamiltonians}
	 f \, \H =\frac{\partial \tilde{S}}{\partial t} - \tilde{P}_a\frac{\partial \tilde{Q}^a}{\partial t} + \mathscr{K}\,.
	\eeq

As in the time-independent case, in order to find the conditions on the generating function $\tilde{S}(q^{a},\tilde{Q}^{a},S,t)$ 
we assume that the coordinates $(q^{a},\tilde{Q}^{a},S,t)$ are independent. 
Thus, from \eqref{canonicalcontact1} one finds that $\tilde{S}$ must satisfy~\eqref{gen-fun} and the additional constraint
	\beq\label{K-to-H}
	f \H= \frac{\partial \tilde{S}}{\partial t} + \mathscr{K}\,,
	\eeq
which defines the new contact Hamiltonian for the new coordinates.  
In the special case $f=1$ the generating function reduces to 
$\tilde S=S-F_1(q^{a},\tilde{Q}^{a},t)$, 
where $F_1(q^{a},\tilde{Q}^{a},t)$ is the generating function of the time-dependent canonical transformation, cf.~\eqref{conditions-Ftime}.

{Now let us consider also the condition on $\tilde S$ imposed by invariance of the  {first} equation in~\eqref{t-dep-dynamics-conditions}.
Rewriting such equation after the transformation we get
 {
	\beq
	\d \tilde \eta^{\tiny\mbox{E}}(\tilde X_{\H}^{\tiny\mbox{E}})=\tilde g_{\mathscr K}\, \tilde\eta^{\tiny\mbox{E}}\,.
	\eeq
with 
	\beq
	\tilde g_{\mathscr K}=-\frac{\partial \mathscr K}{\partial \tilde S}
	\eeq
in the new coordinates, cf.~the first condition in~\eqref{conditions74}.
}
Using that $\tilde \eta^{\tiny\mbox{E}}=f \eta^{\tiny\mbox{E}}$ and that $\tilde X_{\H}^{\tiny\mbox{E}}=X_{\H}^{\tiny\mbox{E}}$ and the two equations in~\eqref{t-dep-dynamics-conditions},
one arrives directly at the following relation
 {
	\beq\label{lastCondonS}
	f\tilde g_{\mathscr K}=f g_{\H}-\d f(X_{\H}^{\tiny\mbox{E}})\,.
	\eeq
}
Notice that for $f=1$, which corresponds to canonical transformations,~\eqref{lastCondonS} reads  {$\tilde g_{\mathscr K}=g_{\H}$}, from which we infer that if
$\H$ does not depend on $S$, then $\mathscr K=0$ is a possible solution of~\eqref{lastCondonS} 
and in such case~\eqref{K-to-H} reduces to the standard Hamilton-Jacobi transformation~\eqref{HJ1}. 
 {However, in the general case $f$ is a function of the extended phase space and thus time-dependent contact transformations extend canonical transformations, as we show
with the following example.}
}

To illustrate the formalism developed so far, we consider an example of an important time-dependent contact transformation, i.e.~we prove
that the Caldirola-Kanai  Hamiltonian~\eqref{CK-H} and the contact Hamiltonian~\eqref{H-lin-S} -- which both give the same damped Newtonian equation -- are related by a time-dependent
contact transformation with $f=e^{\gamma t}$.
To do so, let us consider the Caldirola-Kanai Hamiltonian $H_{\tiny\mbox{CK}}$ as a function on the extended contact phase space $\mathcal T^{\tiny\mbox{E}}$ 
written in the coordinates $(q_{\tiny\mbox{CK}},p_{\tiny\mbox{CK}},S_{\tiny\mbox{CK}},t)$ and the contact Hamiltonian $\H_{S}$ as a function on $\mathcal T^{\tiny\mbox{E}}$ written in the
coordinates $(q,p,S,t)$.
Defining the change of coordinates~\cite{schuch1997nonunitary, schuch2011dynamical, cruz2015time, cruz2016time}
	 \beq\label{CK-to-CH}
	(q,p,S,t) \to (q_{\tiny\mbox{CK}}=q,p_{\tiny\mbox{CK}}=e^{\gamma t}p,{S}_{\tiny\mbox{CK}}=e^{\gamma t}S,t)\, ,
	 \eeq
it is easy to check that the conditions~\eqref{contact1}-\eqref{contact3} and~\eqref{relSHamiltonians} are satisfied
and therefore~\eqref{CK-to-CH} is a time-dependent contact transformation.


\subsection{Hamilton-Jacobi formulation}\label{secHJ}

In this section we introduce a Hamilton-Jacobi formulation of contact Hamiltonian systems. 
This formulation has a major importance, because it establishes a connection with the configuration space, 
where the phenomenological equations are defined. 

The Hamilton-Jacobi equation is a re-formulation of the dynamical equations in terms of a single partial differential equation (PDE)
for the function $S(q^{a},t)$. Thus, we are looking for a PDE of the form 
	\beq\label{Figualacero}
	\mathbb F\left(q^{a},\frac{\partial S}{\partial q^{a}},S,t,\frac{\partial S}{\partial t}\right)=0\,,
	\eeq
whose characteristic curves are equivalent to the contact Hamiltonian dynamics~\eqref{z3}-\eqref{z1}.
To construct such PDE, let us define the function
	\beq\label{ExtendedHJ}
	\mathbb F\left(q^{a},p_{a},S,t,E\right)\equiv E-\H(q^{a},p_{a},S,t)\,.
	\eeq
It turns out that the solution of the equation $\mathbb F=0$ on the configuration space defined by 
	\beq\label{CHJ0}
	\eta^{\tiny\mbox E}=\d S-p_{a}\d q^{a}+\mathscr{H}\d t=0\,,
	\eeq
that is by the two conditions
	\beq\label{CHJ1}
	p_{a}=\frac{\partial S}{\partial q^{a}} \quad {\rm and} \quad \H\left(q^{a},\frac{\partial S}{\partial q^{a}},S,t\right)=-\frac{\partial S}{\partial t}
	\eeq 
gives exactly the contact Hamiltonian equations~\eqref{z3}-\eqref{z1}, together with $\dot t=1$ and $\dot{\H}=-\H\partial \H/\partial S+\partial \H/\partial t$,
which is the evolution of the time-dependent contact Hamiltonian~\eqref{evol-fun-TD}.
Therefore we call the second equation in~\eqref{CHJ1} the \emph{contact Hamilton-Jacobi equation}.

In the symplectic case the Hamilton-Jacobi equation is also the time-dependent canonical transformation induced by the Hamiltonian dynamics.
To find an equivalent formulation for the contact case, 
one must find a generating function $\tilde S(q^{a},Q^{a},S,t)$ 
such that~\eqref{K-to-H} reduces to~\eqref{CHJ1},
where from~\eqref{Lieeta} the function $f$ is 
	\beq
	f=\exp\left(-\int_{0}^{t}\frac{\partial \H}{\partial S}\d \tau\right)\,.
	\eeq
However, contrary to the symplectic case, 
in general such transformation does not lead to a vanishing $\mathscr K$, cf.~equation~\eqref{lastCondonS}.

 {In the~\ref{appendix-B} we give an alternative proof of the equivalence between the contact Hamilton-Jacobi equation
and~\eqref{z3}-\eqref{z1}. 
Such proof is useful because it yields explicitly the algebraic conditions needed to recover the solution $q^{i}(t)$ 
from knowledge of the complete
solution of \eqref{CHJ1}, cf.~equation 
\eqref{conditionbpunto}. An example of such procedure is worked out 
in detail in section \ref{3rdroute}.}

 {
\subsection{Example: the damped parametric oscillator}
In this section we provide an important example, which enables us to show the usefulness of our formalism. The example considered here is
the one-dimensional damped parametric oscillator with mass $m$ and time-dependent frequency $\omega(t)$,
whose contact Hamiltonian is
	\beq \label{D-O}
	\H = \frac{p^2}{2m} +  \frac{1}{2}m\omega^2(t) q^2 + \gamma\,S \,.
	\eeq
Clearly the damped harmonic oscillator is obtained for $\omega(t)=\omega_0$ and the damped free particle is recovered when $\omega(t)=0$. 
The dynamics of the system is given by the contact Hamiltonian equations \eqref{linearq}-\eqref{linearS}, with the time-dependent potential 
$V=\frac{1}{2}m\omega^2(t) q^2$. 
Our aim is to use the tools of contact geometry to solve the dynamics. We show three different ways to solve this system, the first of them using 
contact transformations, the second one using the integrals of motion and the last one by means of the contact Hamilton-Jacobi equation. }

 {
\subsubsection{First route to the solution: contact transformations} \label{1stroute}
In this section we show how to use time-dependent contact transformations to reduce the system to a known form and thus find a solution. 
Let us start by introducing the contact transformation
	\beq\label{Ex-T} 
	(q_{\tiny \mbox{E}},\, p_{\tiny \mbox{E}},\, S_{\tiny \mbox{E}},t) = \left(q\, e^{\frac{\gamma t}{2}},\, \left[
	 p + \frac{m\gamma}{2} q \right]e^{\frac{\gamma t}{2}},\, \left[ S + \frac{m\gamma}{4}q^2 \right]e^{\gamma t},t \right)\,.
	\eeq
The new coordinates $q_{\tiny \mbox{E}}$, $p_{\tiny \mbox{E}}$ and $S_{{\tiny \mbox{E}}}$ are known in the literature 
as the \emph{expanding coordinates}~\cite{schuch2011dynamical, cruz2015time, cruz2016time}. 
The new Hamiltonian in these coordinates is obtained from \eqref{relSHamiltonians} to be
	\beq
	\H_{\tiny\mbox{E}} = e^{\gamma t} \H - \frac{\partial S_{\tiny\mbox{E}}}{\partial t}= \frac{p^2_{\tiny\mbox{E}}}{2m} + \frac{m}{2}\left( \omega^2(t) 
	- \frac{\gamma^2}{4} \right)q^2_{\tiny\mbox{E}}\,.
	\eeq
The Hamiltonian $\H_{\tiny\mbox{E}}$ is known as the \emph{expanding Hamiltonian} and represents a parametric oscillator 
with shifted frequency $\omega^2(t) - \frac{\gamma^2}{4}$. 
This model has been extensively studied since the sixties and there are many methods to obtain the solutions of the equations of motion, 
see for example~\cite{lewis1968class, lewis1969exact, malkin1973linear}. 
It is interesting to note that in the case $\omega(t) = \omega_0$ the Hamiltonian $\H_{\tiny\mbox{E}}$ itself is an invariant of motion.}

 {
\subsubsection{Second route to the solution: the invariants}\label{2ndroute}
As in the standard symplectic theory, an important tool to solve the contact Hamiltonian equations are the invariants (or first integrals) of the system,
which are functions of the (extended) contact phase space that do not vary along the flow, cf. equation~\eqref{evol-fun-TD}. 
In \ref{appendix-A} we prove 
that the damped parametric oscillator possesses the quadratic invariant
	\beq\label{Erm-Inv}
	\mathscr{I}(q,p,t) = \frac{m \, e^{\gamma\, t}}{2}\left[ \left( \alpha(t)\frac{p}{m} - \left[ \dot{\alpha}(t)-\frac{\gamma}{2}\alpha(t)\right] q \right)^2 +\left( \frac{q}{\alpha(t)} \right)^2 \right]\,,
	\eeq  
where the purely time-dependent function $\alpha(t)$ satisfies the \emph{Ermakov equation} 
	\beq\label{Erm-Con}
	\ddot{\alpha} + \left( \omega^2(t) - \frac{\gamma^2}{4} \right) \alpha = \frac{1}{\alpha^3}\, ,
	\eeq
and the S-dependent invariant
	\beq\label{Act-Inv}
	\mathscr{G}(q,p,S,t) = e^{\gamma t} \left[ S - \frac{q\,p}{2}\right]\, .
	\eeq
The invariant $\mathscr I (q,p,t)$ is a generalization of the canonical invariant found by H. R. Lewis Jr. 
for the parametric oscillator~\cite{lewis1968class}, which is recovered when $\gamma \to 0$. 
Besides, the invariant $\mathscr G$ is completely new.
}

 {
To solve the equations of motion of the system~\eqref{D-O} in the general case, 
we use the invariants $\mathscr{I}$ and $\mathscr{G}$ to define the time-dependent contact transformation
	\begin{eqnarray}\label{tran-1}
	\tilde{Q} &=& \arctan\left( \alpha \left[ \dot{\alpha}-\frac{\gamma}{2}\alpha \right] -\alpha^2\frac{p}{m\,q} \right) \\
	\tilde{P} &=&  \mathscr{I}(q,p,t)\\
	\tilde{S}\,&=& \G(q,p,S,t) \label{tran-3}\\
	t\,&=&t\,.
	\end{eqnarray}
The conformal factor in equation \eqref{canonicalcontact1} for this transformation is $f=e^{\gamma t}$ and the new contact Hamiltonian, 
from equation \eqref{relSHamiltonians},  takes the simple form
	\beq 
	\mathscr{K} = \frac{\mathscr{I}}{\alpha^2} \, .
	\eeq
Thus, as $\mathscr{K}$ does not involve the variables $\tilde Q$ and $\tilde{S}$, the new contact Hamiltonian equations have the trivial form
	\beq
	 \dot{\tilde{Q}}^a = \frac{1}{\alpha^2}\, , \quad
	 \dot{\tilde P}_a = 0 \, , \quad
	 \dot {\tilde{S}} = 0 \, , 
	\eeq
with solutions
	\beq \label{sol-triv}
	\tilde Q(t) = \int^t \frac{\d \tau}{\alpha^2(\tau)}\, , \quad 
	\tilde P(t) = \I \quad \mbox{and} \quad 
	\tilde{S}(t) = \G \, .
	\eeq
Now, inverting the transformation \eqref{tran-1}-\eqref{tran-3} and using \eqref{sol-triv}, one obtains the solutions in the original (physical) coordinates, namely
	\begin{eqnarray}
	q(t) &=& \sqrt{\frac{2 \I}{m}}e^{\gamma t} \alpha(t)\cos{\phi(t)}\, ,\label{qt1}\\
	p(t) &=& \sqrt{2\,m \I } \, e^{\gamma t} \left[ \left( \dot{\alpha} - \frac{\gamma}{2}\alpha \right)\cos{\phi(t)} - \frac{1}			{\alpha}\sin{\phi(t)} \right]\, , \label{pt2}\\
	S(t) &=& \mbox{e}^{-\gamma t} \, \G+ \frac{q(t)p(t)}{2} \,\label{St3} ,
	\end{eqnarray}
where $\phi(t)=\tilde{Q}(t)$ and the values of the constants $\I$ and $\G$ are determined by the initial conditions. 
Therefore, we have derived here the solutions of the equations of motion of the damped parametric oscillator
   using the invariants of the contact Hamiltonian system and a proper contact transformation.
From \eqref{qt1}-\eqref{St3} we see that all the dynamics of the system is encoded in the Ermakov equation \eqref{Erm-Con}.
}

 {
\subsubsection{Third route to the solution: the contact Hamilton-Jacobi equation}\label{3rdroute}
We show here another way to find the evolution of the system \eqref{D-O}, that is,
by solving the corresponding contact Hamilton-Jacobi equation \eqref{CHJ1}, which in this case reads
	\beq\label{CHJE}
	 \frac{1}{2m}\left( \frac{\partial S}{\partial q} \right)^2 + \frac{1}{2}m\omega^2(t)q^2+ \gamma S=-\frac{\partial S}{\partial t} \, .
	\eeq	
Due to the form of the left hand side of the above equation, one can propose that $S(q,t)$ is a polynomial with respect to $q$. Thus we 
choose the ansatz
	\beq\label{ansatzS}
	S(q,t) = m\, C(t) \left[ \frac{q^2}{2} - \lambda(t) q \right] + m q \dot{\lambda} + s(t)\,,
	\eeq
where $C(t)$, $\lambda(t)$ and $s(t)$ are purely time-dependent functions. It follows directly that
	\beq
	p(q,t) = \frac{\partial S}{\partial q} = m\, C(t)\left[ q - \lambda(t) \right] + m \dot{\lambda}(t) \,.
	\eeq	
Besides, inserting $S(q,t)$ into the contact Hamilton-Jacobi equation \eqref{CHJE} 
and comparing the coefficients of the same order in $q$, we can find the conditions on $C(t)$, $\lambda(t)$ and $s(t)$. 
After a direct calculation one obtains that $C(t)$ obeys the \emph{Riccati equation}
	\beq\label{R-eq}
	\dot{C} + C^2 + \gamma \, C + \omega^2(t) = 0\, ,
	\eeq  
$\lambda(t)$ satisfies the damped Newtonian equation
	\beq\label{N-eq}
	\ddot{\lambda} + \gamma\,\dot{\lambda} + \omega^2(t) \, \lambda = 0
	\eeq
and
	\beq\label{f-eq}
	\dot{s} = - \frac{m}{2}\left[ C^2 \lambda^2 -2 C\lambda\dot{\lambda} + \dot{\lambda}^2 \right]
	- \gamma s \, .
	\eeq
Now one can use the Riccati equation \eqref{R-eq} and the Newton equation \eqref{N-eq}
to integrate \eqref{f-eq} and obtain
	\beq
	s(t) = \frac{m}{2}\left[ C(t)\lambda^2(t) - \lambda(t)\dot{\lambda}(t) \right]\, .
	\eeq
Substituting into \eqref{ansatzS}, one finds that the solution of the contact Hamilton-Jacobi equation \eqref{CHJE} is
	\beq\label{solutionSqt}
	S(q,t) = \frac{m}{2} C(t)\left[ q - \lambda(t) \right]^2 + m\dot{\lambda}(t) \left[ q-\lambda(t) \right] 
	+ \frac{m}{2} \lambda(t)\dot{\lambda}(t)\, .
	\eeq
Let us mention that solutions $C(t)$ of the Riccati equation are connected to solutions $\lambda(t)$ of the damped 
Newton equation by means of the transformation $C(t)={\dot{\lambda}(t)}/{\lambda(t)}$. 
Therefore, in order to determine $S(q,t)$ it is sufficient to solve only one of these equations.
}

 {Now given the solution \eqref{solutionSqt} of the contact Hamilton-Jacobi equation \eqref{CHJ1},
depending on the non-additive constant $C_{0}=C(0)$,
the trajectory of the particle $q(t)$ can be obtained using \eqref{S-con} and \eqref{conditionbpunto} as follows. 
For this system \eqref{conditionbpunto} implies $b(t) = b_0 \, e^{-\gamma t}$ and therefore \eqref{S-con} reads
	\beq
	\frac{\partial S}{\partial C_0} = \frac{m}{2} \frac{\partial C}{\partial C_0} q^2 = b_0 \, e^{- \gamma t}\,,
	\eeq 
which can be inverted to find the solution trajectory
	\beq\label{tray-q}
	q(t) = \sqrt{\frac{2\,b_0\, e^{- \gamma t}}{m}\left( \frac{\partial C}{\partial C_0} \right)^{-1}}\, .
	\eeq	
For instance, in the case of a damped free particle, $\omega(t) = 0$, 
the solution of the Riccati equation is
	\beq
	C(t) = \frac{ \, e^{- \gamma t}}{\frac{1}{C_0}+ \frac{1}{\gamma}\left( 1- e^{- \gamma t} \right)}\, , 
	\eeq
and from \eqref{tray-q} we can recover the correct trajectories
	\beq
	q(t) = \sqrt{\frac{2 b_0}{m}} \left[ 1 + \frac{C_0}{\gamma} \left( 1 - e^{-\gamma t} \right)\right]\, ,
	\eeq
where the constants $b_0$ and $C_0$ are related to the initial conditions via
	\beq
	b_{0}=\frac{m}{2}q_{0}^{2} \qquad \text{and} \qquad C_{0}=\dot{q}_{0}\,.
	\eeq
} 

 {
Interestingly, with the method presented in this section the evolution of the particle is ultimately determined by the solution of the Riccati equation~\eqref{R-eq},
while with the method given in section \ref{1stroute} one has to directly solve the Newton equation arising from $\H_{\tiny\mbox{E}}$ and 
with the method introduced in section \ref{2ndroute} the solution is given in terms of the solution of the Ermakov equation~\eqref{Erm-Con}.
{This shows that the three methods presented here within the framework of contact geometry are related to the three standard techniques for the solution
of this system.}}


\section{Conclusions and perspectives}
\label{section-conclusions}
In this work we have proposed a new geometric perspective for the Hamiltonian description of mechanical systems. The defining features of our
formulation are that the phase space of any (dissipative or non-dissipative) mechanical system is assumed to be a contact manifold 
and that the evolution equations are given as contact Hamiltonian equations, see~\eqref{z3}-\eqref{z1}.
We have shown that contact Hamiltonian dynamics on the one hand recovers all the results of standard symplectic dynamics when the contact Hamiltonian
$\H$ does not depend explicitly on $S$
and on the other hand
can account for the evolution of systems with different types of dissipation in the more general case in which $\H$ depends on $S$. 

We have considered both time-independent and time-dependent contact systems and we have found in both cases the transformations (called contact transformations) 
that leave the contact Hamiltonian equations invariant,
showing that canonical transformations of symplectic dynamics are a special case. To show the usefulness of contact transformations, 
we have provided an explicit example (the Caldirola-Kanai model for systems with linear dissipation) in which
a non-canonical but contact transformation~\eqref{CK-to-CH} allows to move from the usual time-dependent canonical description in terms 
of non-physical variables to a contact description in terms of the physical variables.

By computing the divergence of the contact Hamiltonian flow~\eqref{divergence}, we have provided a formal definition 
of dissipation in our formalism in terms of the contraction of the phase space, which is usually associated with
irreversible entropy production~\cite{daems1999entropy,gallavotti2004nonequilibrium}. 

 {In addition}, we have derived a contact Hamilton-Jacobi equation~\eqref{CHJ1} 
whose complete solution is equivalent to solve the Hamiltonian dynamics, as it happens in standard symplectic mechanics.

 {Finally, we have worked out in detail a specific important example (the damped parametric oscillator)
for which we have solved the  dynamics in three different ways: using contact transformations, using the invariants of the system
and resorting to the solution of the associated contact Hamilton-Jacobi equation.
This example thus provides a direct evidence of the usefulness of our formalism.}

Given the importance of the symplectic perspective in the classical mechanics of conservative systems,
we consider that the contact perspective could play a similar role in the mechanics of dissipative systems.
For instance, a relevant question is that of a quantization of our formalism. Here we sketch briefly such possibility.
 Using the fact that the additional contact variable is a generalization of Hamilton's principal function which satisfies the contact version of the Hamilton-Jacobi equation,  
and that the canonical momenta and positions in our formalism coincide with the physical ones,
we suggest a canonical quantization of the contact Hamiltonian based on the standard rules of canonical quantization, namely 
	\beq\label{quantization}
	p_{a}\rightarrow \frac{\hbar}{i}\frac{\partial}{\partial q^{a}},\quad q^{a}\rightarrow \hat{q}^{a},\quad S(q^{a},t)\rightarrow \frac{\hbar}{i}\,{\rm ln}\,\Psi(q^{a},t)\,.
	\eeq
Using such rules to quantize the contact Hamiltonian $\H$ and obtain the operator $\hat{\H}$, one can define the 
 {``contact Schr\"odinger equation''}
	\beq\label{tdepSE}
	i\hbar \frac{\partial \Psi}{\partial t}=\hat{\H} \Psi\,.
	\eeq
This equation has a fundamental property: in the case in which the contact Hamiltonian reduces to a symplectic Hamiltonian (i.e.~when $\H$ does not depend on $S$
explicitly) and the dynamics reduces to a standard conservative dynamics, equation~\eqref{tdepSE} obviously reduces to the standard linear Schr\"odinger equation
and all the known results for the quantization of conservative systems are recovered. 
However, this equation has the disadvantage that in general it does not conserve the norm of the wave function~\cite{schuch2011dynamical}. For systems with contact Hamiltonian of the form $\H = H_{\tiny \mbox{mec}}+h(S)$, see~\eqref{H-diss},
normalization is achieved following the procedure of Gisin~\cite{gisin1983irreversible},
which consists in subtracting the mean value of $h$, that is $\langle \hat{h} \rangle = \int \Psi^\ast \, \hat{h} \, \Psi\, \d q^a$.
This leads to the nonlinear Schr\"odinger equation
	\beq\label{G-Q}
	i\hbar \frac{\partial \Psi}{\partial t}=\hat{H}_{\tiny\mbox{mec}}\Psi + (\hat{h} - \langle \hat{h} \rangle)\Psi\,.
	\eeq
Applying~\eqref{G-Q} to
the contact Hamiltonian $\H_{S}$ with linear dependence on $S$ given in equation~\eqref{H-lin-S} and using~\eqref{tdepSE}, we get the evolution equation
	\beq\label{NLSE}
	i\hbar \frac{\partial \Psi}{\partial t}=\left(-\frac{\hbar^{2}}{2m}\nabla^{2}+V(q^{a})+\gamma \frac{\hbar}{i}\left[{\rm ln}\,\Psi	-\langle {\rm ln} \Psi \rangle \right]\right)\Psi\,,
	\eeq	
which is exactly the phenomenological nonlinear Schr\"odinger equation introduced 
in~\cite{schuch1983nonlinear, schuch1984nonlinear, schuch1986macroscopic} for the description of dissipative systems, 
see also~\cite{schuch1997nonunitary, schuch2011dynamical, cruz2015time, cruz2016time}. 

This fact, together with the result that the contact dynamics generated by
$\H_{S}$ coincides with the classical Newtonian equations for systems with linear dissipation~\eqref{New-CK},
provides a further theoretical justification for the introduction of the nonlinear phenomenological Schr\"odinger equation~\eqref{NLSE}
and, most importantly, displays an intriguing consistency between the classical and quantum descriptions in our proposal.
A more detailed study of the extension of contact Hamiltonian mechanics to quantum systems will be presented in a future work.
 {For instance, it will be worth trying a more geometric quantization program, e.g.~following the lines of \cite{fitzpatrick2011geometric}.}

Finally, we have not considered here the Lagrangian formulation. This aspect is fundamental in order to have a complete picture of contact mechanics
and it is of primary interest for extension to field theory.

\section*{Acknowledgements}
The authors would like to thank the unknown referee for interesting comments and suggestions.
AB is funded by a DGAPA postdoctoral fellowship. DT acknowledges financial support from CONACyT, CVU No. 442828.


\appendix

 {
\section{Invariants for the damped parametric oscillator}
\label{appendix-A}
In this appendix we prove that $\mathscr{I}(q,p,t)$ and $\mathscr{G}(q,p,S,t)$ given in equations \eqref{Erm-Inv} and \eqref{Act-Inv} are two invariants of the damped
parametric oscillator defined by the contact Hamiltonian \eqref{D-O}.
An invariant is a function $\mathscr F$ of the (extended)
 contact phase space that satisfies the partial differential equation
	\beq\label{PDE}
	 -\mathscr{H}\,\frac{\partial \mathscr{F}}{\partial S}
	+p_a\left\{\mathscr{F},  \mathscr{H}\right\}_{(S,p_a)}
	+ \left\{\mathscr{F},  \mathscr{H}\right\}_{(q^a,p_a)}
	= - \frac{\partial \mathscr{F}}{\partial t}\,,
	\eeq 
where we use the same notation as in \eqref{ContactEvolution}.
To find a solution, we propose the ansatz 
	\beq\label{inv-1}
	\mathscr{F}(q,p,S,t) = \beta(t)p^2 - 2\xi(t) qp + \eta(t) q^2 + \zeta(t) S \,.
	\eeq
Inserting \eqref{inv-1} into \eqref{PDE}, we get the system of ordinary differential equations  
	\begin{eqnarray}
	\dot{\beta} &=&\frac{2}{m}\xi + 2\gamma \beta - \frac{1}{2m} \zeta \, , \label{A3}\\
	\dot{\eta} &=& -2 m\omega^2\xi + \frac{1}{2}m\omega^2\zeta\, , \\
	\dot{\xi} &=& \frac{1}{m}\eta+\gamma\xi - m\omega^2\beta \, , \label{A5}\\
	\dot{\zeta} &=& \gamma\,\zeta \, .
	\end{eqnarray}
Then clearly 
	\beq\label{solzeta}
	\zeta(t)=\zeta_0e^{\gamma t}\,,
	\eeq 
and we are left with the problem of solving the system \eqref{A3}-\eqref{A5}. 
To do so, we consider the change of variables $\tilde{\beta}(t) = e^{-\gamma t} \beta(t)$, $\tilde{\eta}(t) = e^{-\gamma t} \eta(t)$ and $\tilde{\xi}(t) = e^{-\gamma t} \xi(t)$, 
which yields the equivalent system
	\begin{eqnarray}
	\dot{\tilde{\beta}} &=& \frac{2}{m} \tilde{\xi} +\gamma \tilde{\beta} - \frac{\zeta_0}{2m}\, , \\
	\dot{\tilde{\eta}} &=& -2m\omega\tilde{\xi} - \gamma \tilde{\eta} + \frac{\zeta_0}{2}m\omega^2\, ,\\
	\dot{\tilde{\xi}} &=& \frac{1}{m}\tilde{\eta} - m\omega^2\tilde{\beta}\, .
	\end{eqnarray}
To solve this system, we re-write it as a third-order ordinary differential equation for $\tilde{\beta}(t)$
	\beq\label{t-ord-eq}
	\dddot{\tilde{\beta}} + 4 \, \Omega^2 \, \dot{\tilde{\beta}} + 4 \Omega\, \dot{\Omega}\,\tilde{\beta} = 0\,,
	\eeq
where for simplicity we define $\Omega^2 = \omega^2 - \frac{\gamma^2}{4}$. 
The above equation is known as {\emph{the normal form} of a third order equation of maximal symmetry}~\cite{leach2008ermakov}.
}

 {
Now, using the further change of variable 
	\beq \label{solbeta}
	\tilde{\beta}(t) = \frac{1}{2m} \alpha^2(t)
	\eeq
in \eqref{t-ord-eq}, one obtains that $\tilde\beta(t)$ is a solution of \eqref{t-ord-eq} if and only if $\alpha(t)$ is a solution of the Ermakov equation \eqref{Erm-Con}. 
Moreover, from \eqref{solbeta} one can re-write the remaining two equations as 
	\begin{eqnarray}
	\tilde{\eta}(t) &=& \frac{m}{2} \left( \left[ \dot{\alpha}(t) - \frac{\gamma}{2}\alpha(t) \right]^2 + \frac{1}{\alpha^2(t)} \right)\, ,\label{soleta} \\
	\tilde{\xi}(t) &=& \frac{\alpha(t)}{2}\left( \dot{\alpha}(t) - \frac{\gamma}{2}\alpha(t) \right)  + \frac{1}{4}\,\label{solxi} .
	\end{eqnarray}
}

 {
Finally, using \eqref{solzeta},\eqref{solbeta}-\eqref{solxi} and $\beta(t) = e^{\gamma t} \tilde{\beta}(t)$, $\eta(t) = e^{\gamma t} \tilde{\eta}(t)$, $\xi(t) = e^{\gamma t} \tilde{\xi}(t)$ 
into the ansatz \eqref{inv-1}, we find that 
	\beq
	\mathscr{F}(q,p,S,t)=\mathscr{I}(q,p,t)+\zeta_{0}\mathscr{G}(q,p,S,t)\,,
	\eeq
with
	\beq
	\mathscr{I}(q,p,t) = \frac{m \, e^{\gamma\, t}}{2}\left[ \left( \alpha(t)\frac{p}{m} - \left[ \dot{\alpha}(t)-\frac{\gamma}{2}\alpha(t)\right] q \right)^2 +\left( \frac{q}{\alpha(t)} \right)^2 \right]\, ,
	\eeq  
and
	\beq
	\mathscr{G}(q,p,S,t) = e^{\gamma t} \left[ S - \frac{q(t)p(t)}{2} \right]\,.
	\eeq
Since $\mathscr{F}(q,p,S,t)$ is an invariant for any choice of the initial conditions and since $\zeta_{0}$ only depends on the initial conditions, it follows that
$\mathscr{I}(q,p,t)$ and $\mathscr{G}(q,p,S,t)$ separately are invariants of the system.
}


 {
\section{Equivalence between the contact Hamilton-Jacobi equation and the contact Hamiltonian equations}
\label{appendix-B}
In this appendix we prove that finding the complete solution of the contact Hamilton-Jacobi equation~\eqref{CHJ1} 
is equivalent to solving the equations of motion \eqref{z3}-\eqref{z1}. This proof is a generalization of the standard proof for the symplectic case~\cite{goldstein2002classical}.
}

 {
To begin, let $S(q^1,\dots,q^n, c^1,\dots,c^{n},t)$ be the complete solution of \eqref{CHJ1}, where $c^i$ are $n$ constants and suppose 
	\begin{equation}
	\left| \frac{\partial^2 S}{\partial q^i \partial c^j} \right| \neq 0\,.
	\end{equation}
Using the quantities $p_i(q^1,\dots,q^n, c^1,\dots,c^{n},t)=\frac{\partial S}{\partial q^{i}}$, we can rewrite  \eqref{CHJ1} as
	\beq\label{Contact-Hamiltonian-HJ}
	\mathscr{H}\left(q^{1},\dots,q^{n},p_1,\dots,p_{n},S,t\right)=-\frac{\partial S}{\partial t}\,.
	\eeq
Besides, defining
	\beq \label{S-con}
	b_i=\frac{\partial S}{\partial c^i},
	\eeq 
we obtain
	\begin{equation}\label{bidot}
	\dot{b}_i = \frac{\partial^2 S}{\partial q^j \partial c^i}\dot{q}^j+\frac{\partial^2 S}{\partial t\,\partial c^i}
	\end{equation}
and deriving \eqref{Contact-Hamiltonian-HJ} with respect to $c^i$ we have 
	\begin{equation}\label{secondder1}
	\frac{\partial^2 S}{\partial c^i \partial t}=-\frac{\partial\mathscr{H}}{\partial S}b_i-\frac{\partial\mathscr{H}}{\partial p_j}\frac{\partial^2 S}{\partial c^i\partial q^j}\,.
	\end{equation}
Combining \eqref{bidot} and \eqref{secondder1} we get
	\begin{equation}\label{b}
	\dot{b}^i = \frac{\partial^2 S}{\partial q^j \partial c^i}\left( \dot{q}^j -\frac{\partial \mathscr{H}}{\partial p_j} \right)-\frac{\partial \mathscr{H}}{\partial S}b_i\,.
	\end{equation} 
Now, from the definition of $p_i$, it follows that
	\begin{equation}\label{momentum}
	\dot{p}_i = \frac{\partial^2 S}{\partial q^j \partial q^i}\dot{q}^j+\frac{\partial^2 S}{\partial t\, \partial q^i}\,,
	\end{equation}
and deriving \eqref{Contact-Hamiltonian-HJ} with respect to $q^i$ one obtains
	\begin{equation}\label{Ham}
	\frac{\partial^2 S}{\partial q^i\partial t}=-\frac{\partial\mathscr{H}}{\partial q^i}-\frac{\partial\mathscr{H}}{\partial S}p_i-\frac{\partial\mathscr{H}}{\partial p_j}\frac{\partial^2 S}{\partial q^i \partial q^j}\,.
	\end{equation}
From \eqref{momentum} and \eqref{Ham} we get
	\begin{equation}\label{p}
	\dot{p}_i=\frac{\partial^2 S}{\partial q^j \partial q^i}\left( \dot{q}^j -\frac{\partial \mathscr{H}}{\partial p_j} \right)-\frac{\partial \H}{\partial q^i}-p_{i}\frac{\partial \H}{\partial S}\,.
	\end{equation}
It is thus easy to see that imposing 
	\beq\label{conditionbpunto}
	\dot{b}_i =-\frac{\partial \H}{\partial S}b_i\,,
	\eeq 
equations \eqref{b} and \eqref{p} reduce to
	\begin{eqnarray}
	\dot{q}^{i} &=& \frac{\partial \mathscr{H} }{\partial p_i} \,,\label{qidot}\\
	\dot{p}_{i} &=&  -\frac{\partial \mathscr{H}}{\partial q^i} - p_i \frac{\partial \mathscr{H}}{\partial S}\,,\label{pidot}
	\end{eqnarray}
which coincide with \eqref{z3} and \eqref{z2}. Finally, using the fact that $c^{i}$ are constants of motion,
the equation for the evolution of $S$ reads 
	\begin{equation}\label{Sdot}
	\dot{S}=\frac{\partial S}{\partial q^i}\dot{q}^i+\frac{\partial S}{\partial t}=p_i\,\frac{\partial \mathscr{H} }{\partial p_i}-\mathscr{H}\, ,
	\end{equation}
where in the last equality we have used that $p_{i}=\partial S/\partial q^{i}$, $\dot{q}^{i}=\partial \H/\partial p_{i}$ and that $\H=-\partial S/\partial t$.  
Equations \eqref{qidot}-\eqref{Sdot} are exactly equivalent to \eqref{z3}-\eqref{z1}. 
Therefore we have proved that the contact Hamilton-Jacobi equation \eqref{CHJ1} 
is equivalent to the contact Hamiltonian dynamics, provided the condition \eqref{conditionbpunto} holds. 
Therefore such condition has a primary importance. In fact, this yields the algebraic conditions to be solved for $q^{i}$ in order to recover the solution
$q^{i}(t)$ from knowledge of $S(q^{i},c^{i},t)$.
For an explicit example see section \ref{3rdroute}.
}

\bibliographystyle{ieeetr}
\bibliography{Conservative}

\end{document}